\documentclass[dvipsnames,aps,prl,nobibnotes,twocolumn,superscriptaddress,nolongbibliography]{revtex4-2}
\setcitestyle{super}
\usepackage[T1]{fontenc}
\usepackage[utf8]{inputenc}

\usepackage[unicode=true,pdfusetitle,
 bookmarks=true,bookmarksnumbered=false,bookmarksopen=false,
 breaklinks=false,pdfborder={0 0 0},pdfborderstyle={},backref=false,colorlinks=true]{hyperref}
 \hypersetup{
 citecolor=blue,
 urlcolor=blue,
 linkcolor=blue}
\usepackage[table,svgnames]{xcolor}
\usepackage{array}
\usepackage{units}
\usepackage{amsmath}
\usepackage{amssymb}
\usepackage{graphicx}
\usepackage{stackrel}
\usepackage{graphicx}
\usepackage{babel}
\usepackage[final]{pdfpages}
\usepackage{mathtools}
\usepackage{float}
\usepackage{pdfpages}

\makeatletter
\AtBeginDocument{\let\LS@rot\@undefined}
\makeatother

\makeatletter
\def\@hangfrom@section#1#2#3{\@hangfrom{#1#2}#3}%\MakeTextUppercase{#3}}%
\def\@hangfroms@section#1#2{#1#2}%\MakeTextUppercase{#2}}%
\makeatother

\begin{document}

\title{Structural coarse-graining enables noise-robust functional connectivity and reveals hidden inter-subject variability}
%Renormalization group uncovers hidden functional heterogeneity in human brain dynamics

%Uncovering Hidden Population Variability in Brain Networks with Limited fMRI Data

%Heterogeneous Coarse-Graining Reveals Emergent Function in Brain Networks with Limited Temporal Data
\author{Izaro Fernandez-Iriondo}
\email[]{fernandeziriondo.izaro@gmail.com}

\affiliation{Computer Science and Artificial Intelligence, University of the Basque Country (UPV/EHU), San Sebastian, Spain}
\affiliation{Computational Neuroimaging Lab, Biobizkaia Health Research Institute, Barakaldo, Spain}
\affiliation{Informatics Engineering Doctoral Programme, University of the Basque Country (UPV/EHU), San Sebastian, Spain}

\author{Antonio Jimenez-Marin}
\affiliation{Computational Neuroimaging Lab, Biobizkaia Health Research Institute, Barakaldo, Spain}

\author{Jesus Cortes}
\affiliation{Computational Neuroimaging Lab, Biobizkaia Health Research Institute, Barakaldo, Spain}
\affiliation{IKERBASQUE: The Basque Foundation for Science, Bilbao, Spain}
\affiliation{Department of Cell Biology and Histology, University of the Basque Country (UPV/EHU), Leioa, Spain}

\author{Pablo Villegas}
\email[]{pablo.villegas@cref.it}
\affiliation{`Enrico Fermi' Research Center (CREF), Via Panisperna 89A, 00184 - Rome, Italy}
\affiliation{Instituto Carlos I de F\'isica Te\'orica y Computacional, Univ. de Granada, E-18071, Granada, Spain.}

\begin{abstract}
Functional connectivity estimates are highly sensitive to analysis choices and can be dominated by noise when the number of sampled time points is small relative to network dimensionality. This issue is particularly acute in fMRI, where scan resolution is limited. Because scan duration is constrained by practical factors (e.g., motion and fatigue), many datasets remain statistically underpowered for high-dimensional correlation estimation. We introduce a framework that combines diffusion-based structural coarse-graining with spectral noise filtering to recover statistically reliable functional networks from temporally limited data. The method reduces network dimensionality by grouping regions according to diffusion-defined communication. This produces coarse-grained networks with dimensions compatible with available time points, enabling random matrix filtering of noise-dominated modes. We benchmark three common FC pipelines against our approach. We find that raw-signal correlations are strongly influenced by non-stationary fluctuations that can reduce apparent inter-subject variability under limited sampling conditions. In contrast, our pipeline reveals a broader, multimodal landscape of inter-subject variability. These large-scale organization patterns are largely obscured by standard pipelines. Together, these results provide a practical route to reliable functional networks under realistic sampling constraints. This strategy helps separate noise-driven artifacts from reproducible patterns of human brain variability.
\end{abstract}
%The construction of functional connectivity maps in neuroimaging is highly sensitive to methodological choices, such as the selection of a brain parcellation. Limited temporal data worsens this scenario.
% Neural measurements are often constrained to limited temporal resolution.
%This limitation, usually caused by participant fatigue constraints, motion artifacts, and the cost of scanning time, cannot be remedied by simply collecting more samples, leaving many experimental datasets statistically unreliable.
%To address this problem, we use a heterogeneous coarse-graining strategy based on the recently introduced Laplacian Renormalization Group.
%This approach reduces the dimensionality of high-resolution structural networks while preserving their key diffusion properties, creating coarse-grained regions that meet the statistical requirements for rigorous spectral analysis and allow random noise to be filtered out, yielding significant functional networks.
%Our results address different computational choices either using raw signals, temporal derivatives, or spectrally filtered derivative-based matrices, showing distinct lenses that highlight different aspects of the brain’s function and its relationship to the underlying anatomy.
%Overall, our work provides a robust strategy for extracting reliable networks from temporally limited data, and a clear demonstration that methodological decisions themselves critically shape the structural–functional patterns that can be observed in the human brain.

\maketitle

%\section{INTRODUCTION}
Complex cognitive functions---such as perception, memory, decision-making, and navigation---emerge from the coordinated activity of large neuronal populations \cite{BuzsakiBook,Panzeri2022}. These populations not only encode sensory information but also transmit and integrate it across brain regions to generate appropriate coordinated behavioral responses \cite{Katz2016,Panzeri2015}. Remarkably, even in the absence of external stimuli or explicit tasks, the brain dynamics remains spontaneously active \cite{Softky1993,Arieli1996,Raichle2011}, revealing an interplay in which structural architecture constrains and biases functional interactions without fully determining them \cite{Fotiadis2024,Preti2019,diez_novel_2015}. This intrinsic, non-trivial activity appears to be a fundamental feature of neural computation. Understanding the origin and functional role of this energy-demanding resting --or ``un-resting''-- state, and how it interacts with input-driven responses, remains a central challenge in neuroscience \cite{Fotiadis2024, raichle_brains_2015, hj_structural_2013}, with implications for how the brain processes and transmits information.

Correlation structure provides a direct window onto functional interactions. By revealing patterns of functional connectivity, correlation analysis facilitated the inference of circuit organization in systems ranging from the retina \cite{Greschner2011} to the visual thalamocortical pathway \cite{Reid1995} and local cortical networks \cite{Aertsen1989,Alonso1998}. Changes in correlation structure across stimuli or behavioral states can expose computations that single-neuron activity alone would miss \cite{Ahissar1992,Gutnisky2008,Komiyama2010, cortes_effect_2012}. For example, during active exploration, sensory cortical responses become desynchronized even without changes in input or firing rate \cite{Poulet2008}.  At a larger scale, Functional Magnetic Resonance Imaging (fMRI) enables monitoring of population dynamics across distributed brain regions, providing complementary insight into large-scale functional connectivity. Functional connectivity patterns have also been shown to contain reliable subject-specific signatures across sessions, enabling individual identification from connectome structure \cite{Finn2015,Gratton2018, jo_subject_2021}.

In practice, estimating population-level correlations is limited by sampling and noise \cite{kohn_correlations_2016, kanitscheider_origin_2015, pouget_inference_2003}. Importantly, a substantial fraction of functional correlations in resting-state fMRI is known to be driven by global, non-neuronal fluctuations related to motion, physiology, and arousal \cite{Power2012,Caballero2017}. The treatment of these fluctuations remains debated, particularly in the context of global signal regression, which can both reduce artifacts and alter correlation structure \cite{Murphy2017}. Rather than regressing out global components a priori, our approach targets noise-dominated modes at the spectral level, allowing their statistical identification without imposing a specific preprocessing model. In high-dimensional datasets, noise can obscure genuine interactions, induce spurious correlations, and cause standard null models to misrepresent the true structure of correlation matrices \cite{Cohen2011,MarPas,PRX-Gar}. However, fMRI rarely provides the temporal resolution needed for high-dimensional correlation estimation, making it difficult to separate meaningful structure from sampling noise. Reliable covariance estimation typically requires $T \gtrsim N$ (up to constant factors depending on the estimator and temporal autocorrelation). This limitation is not merely theoretical: empirical studies show that the reliability of functional connectivity estimates depends strongly on scan duration, with shorter acquisitions producing less stable connectivity structure \cite{Birn2013,Noble2019}. Many approaches mitigate noise (e.g., thresholding, sparsity constraints, or regularization), but they necessarily impose assumptions on network structure. Here, we instead use spectral criteria to remove noise-dominated components without assuming sparsity or a specific architecture.

%An appealing framework to understand the complex neural population dynamics at the mesoscale is the criticality hypothesis \cite{Beggs2003,Shew2013,diSanto2018}, which proposes that neural networks operate near a phase transition between distinct dynamical regimes \cite{RMP-MAM,Plenz2014}. High-throughput recordings suggest that this critical behavior may enhance information representation, resulting in scale-invariant activity patterns across neuronal populations \cite{Mora2011,Bellay2015,Cambra2025,Morales2023}. However, apparent signatures of criticality can also arise from latent inputs, temporal subsampling, or noise, especially in high-dimensional datasets, such as fMRI \cite{Calvo2025, Levina2017, Priesemann2009}. Distinguishing genuine critical dynamics from these confounding effects remains challenging and requires analytical approaches that robustly capture bona fide population-level interactions.

A complementary approach is to reduce dimensionality before estimating correlations. The Laplacian Renormalization Group (LRG) reduces network dimensionality while preserving diffusion-defined communication structure \cite{LRG, InfoCore, jstat_statphys, Gabrielli2025}. In neuroimaging, LRG acts as a data-driven alternative to fixed atlases by grouping regions according to structural communication pathways, enabling subject-specific multiscale supernodes; here, however, we focus on a population-level coarse-graining. This coarse-graining yields networks whose size matches the available sampling, enabling more reliable functional connectivity estimates. This allows functional connectivity to be estimated at a scale supported by the data, rather than being imposed a priori.

We introduce a two-step framework—structural coarse-graining followed by spectral noise filtering—that makes functional connectivity estimation statistically well-posed under realistic scan durations. Following validation on synthetic benchmarks, we apply the method to human connectome data, demonstrating that it preserves long-range functional structure often lost in standard analyses. This capability addresses a persistent challenge in human neuroimaging: the accurate estimation of brain organization at the population level under realistic scan durations. By maximizing statistical stability without requiring impractical acquisition times, our approach offers a solution to a key bottleneck in translational and population neuroscience.

\section{RESULTS}
\subsection{De-noising Functional Brain Networks.}
De-noising functional brain networks is challenging because correlations must be estimated in high dimensions from low-resolution recordings. Each time series can be written as
$X_i \equiv \{ x_i(1), x_i(2), \dots, x_i(T) \},$
where \(T\) denotes the number of temporally ordered measurements of the activity of the \(i^\text{th}\) unit. In neuroimaging studies, the index \(i\) typically refers to a specific brain element—for example, a voxel in functional MRI (fMRI), or a sensor in electroencephalography (EEG) or magnetoencephalography (MEG). Thus, \(X_i\) represents the time series of neural activity recorded from that region or sensor across the \(T\) sampled time points.

To quantify the level of dependency between different system units, correlation matrix analysis provides a canonical representation for revealing functional interactions in brain networks. Specifically, the Pearson correlation matrix is defined as 
\[
    C_{ij} \equiv \frac{\mathrm{Cov}[X_i,X_j]}{\sqrt{\mathrm{Var}[X_i] \cdot \mathrm{Var}[X_j]}}.
\]

%\textcolor{red}{meter citas de spurious correlations}

Although various methods have been proposed to filter out spurious correlations, traditional approaches in neuroscience often rely on thresholding techniques, where weak correlations below a fixed cutoff are removed \cite{rubinov_complex_2010}. This strategy can be problematic, as it may discard relevant connections while retaining strong but potentially misleading ones. Alternative approaches include topology-based procedures, such as minimum spanning trees \cite{tewarie_minimum_2015}, which preserve only the strongest links necessary to maintain global connectivity, and statistical inference methods that generate surrogate datasets or use bootstrapping to estimate the likelihood that the observed correlation structure arises by chance \cite{zalesky_use_2012}. While these strategies can reduce noise, they either introduce arbitrary thresholds or require extensive computation and assumptions about the null model, potentially obscuring the underlying functional network configurations. A successful alternative relies on spectral filtering \cite{PRX-Gar,MarPas}. This strategy is grounded in random matrix theory, and allows for isolating meaningful correlations from random noise without thresholding, thereby providing a clearer and more reliable representation of functional brain networks.

% Therefore, the accuracy provided by spectral filtering here deserves our special attention.

For uncorrelated signals, random matrix theory predicts a characteristic eigenvalue spectrum (Marchenko–Pastur), which we use as a noise reference \cite{MarPas},
\begin{equation}
\rho(\lambda)=\frac{T}{N}\,\frac{\sqrt{(\lambda_{+}-\lambda)(\lambda-\lambda_{-})}}{2\pi\,\lambda},
\qquad \lambda_{-}\le \lambda \le \lambda_{+},
\label{MPastur}
\end{equation}
where \(\lambda_\pm = \left[ 1 \pm \sqrt{\frac{N}{T}} \right]^2\), under the asymptotic condition \(N \to \infty\), \(T \to \infty\) with \(1 < \frac{T}{N} < \infty\). The set of eigenvalues described by this distribution can generally be considered as arising from random noise and may therefore be filtered out to isolate statistically significant correlations from any set of temporal recordings \cite{PRX-Gar}. 

Before applying our method to empirical data, we first validated its core components on two well-understood synthetic benchmarks. These benchmarks were chosen to test their ability to solve two distinct problems: (1) spurious correlations arising from non-stationary, non-interacting signals (the Random Walk model), and (2) the separation of ``true'' emergent correlations from noise in a complex, interacting system (the Kuramoto model).

\subsubsection{Non-interacting nonstationary signals}
As a first benchmark to characterize noise-induced correlations, we consider a minimal null model in which no true interactions are present. We generate \( N \) independent time series from a stochastic real variable \( x(t) \) evolving according to a Wiener process—namely, a continuous-time unbiased random walk (RW). The dynamics \cite{Gardiner} is defined by the Langevin equation
\begin{equation}
\dot{x}(t) = \sigma\,\eta(t),
\end{equation}
where \( \eta(t) \) denotes a Gaussian white noise with zero mean and unit variance, and \( \sigma \) sets the noise amplitude. This process represents the simplest artificial null model, providing a baseline for testing and filtering out trivial correlations.

\begin{figure}[hbtp]
\centering
 \includegraphics[width=1.0\columnwidth]{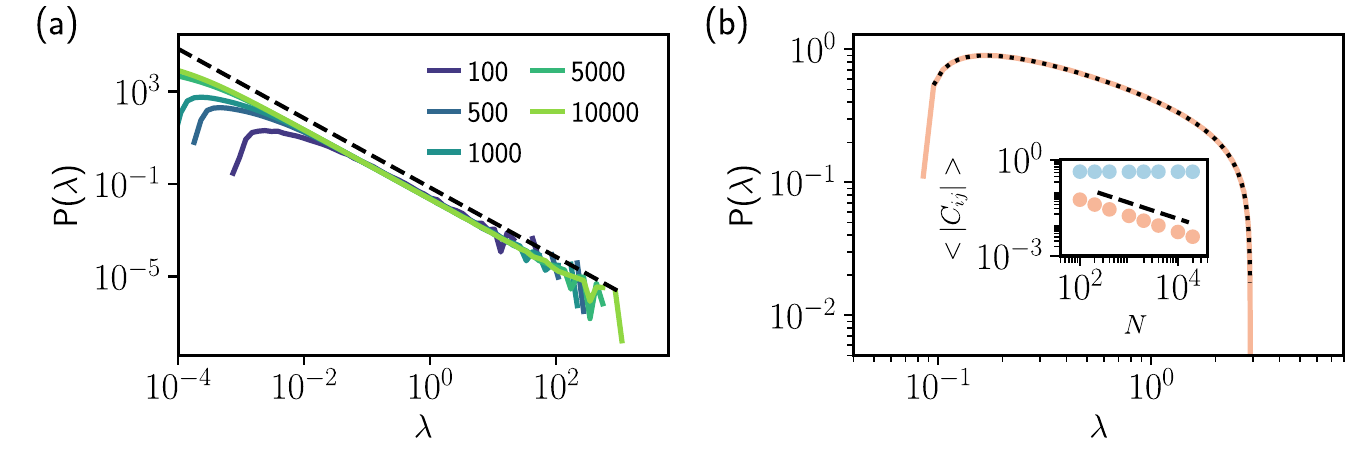}
 \caption{\textbf{Spectral noise analysis of random signals.} (a) Eigenvalue distribution of the Pearson correlation matrix for $N$ non-interacting random walks (see legend for system size). Black dashed line indicates $P(\lambda)\sim\lambda^{-1.5}$.  
(b) Eigenvalue distribution for the Pearson correlation matrix of the derivative of the dynamical evolution for $N=10^4$ random walks. The dashed line shows the expected Marchenko-Pastur distribution. \textit{Inset:} Mean off-diagonal absolute Pearson correlation versus system size for both cases; dashed line shows $|C_{ij}|\sim1/\sqrt{N}$, as expected from the central limit theorem. All curves are averaged over $10^2$ realizations.}\label{RWalk}
\end{figure}

Figure~\ref{RWalk} illustrates that, for the case of purely random temporal series, the Marchenko--Pastur distribution is recovered when the derivative of the stochastic differential equation is analyzed (see Fig.~\ref{RWalk}b). In contrast, when the original non-stationary series \(x(t)\) is considered, the eigenvalue distribution exhibits a power-law tail, \(P(\lambda) \sim \lambda^{-3/2}\) (see Fig.~\ref{RWalk}a and Supplementary Information 1 (SI1) for a formal derivation), as the Marchenko--Pastur prediction does not hold due to the strong temporal autocorrelations inherent to random walks \cite{Calvo2025}.

The inset of Fig.~\ref{RWalk}(b) highlights a \emph{scaling effect}: the mean off-diagonal Pearson correlation between different random walks decreases very slowly as the number of independent series \(N\) increases. Although the processes are independent, the cumulative fluctuations in each random walk generate correlations that decay more slowly than the \(1/\sqrt{N}\) behavior expected for stationary signals. Consequently, even large finite-length datasets can exhibit measurable cross-correlations despite the absence of any interaction. This phenomenon is known as \emph{spurious regression} \cite{BouchaudBook,Granger1974}, which arises when two non-stationary time series exhibit a high statistical correlation that is not due to a genuine causal link, but rather to the accumulation of random variations over time. Interpreting such apparent associations as evidence of causal influence exemplifies a logical error that can easily arise in functional brain studies. Here, temporal derivatives are introduced not as a physiological signal, but as a statistical diagnostic that restores the validity of random matrix predictions in non-stationary regimes. To address this issue, we simply treat \(X_i\) as the temporal derivative of the neural recording, enabling a proper detection of white-noise effects in the spectral domain. Importantly, temporal differentiation acts as a whitening step even for stationary but autocorrelated signals, as confirmed by Ornstein–Uhlenbeck controls (see SI2).

\subsubsection{Emergence of Non-Trivial Correlations in Network Dynamics}  

\begin{figure}[hbtp]
\centering
 \includegraphics[width=0.95\columnwidth]{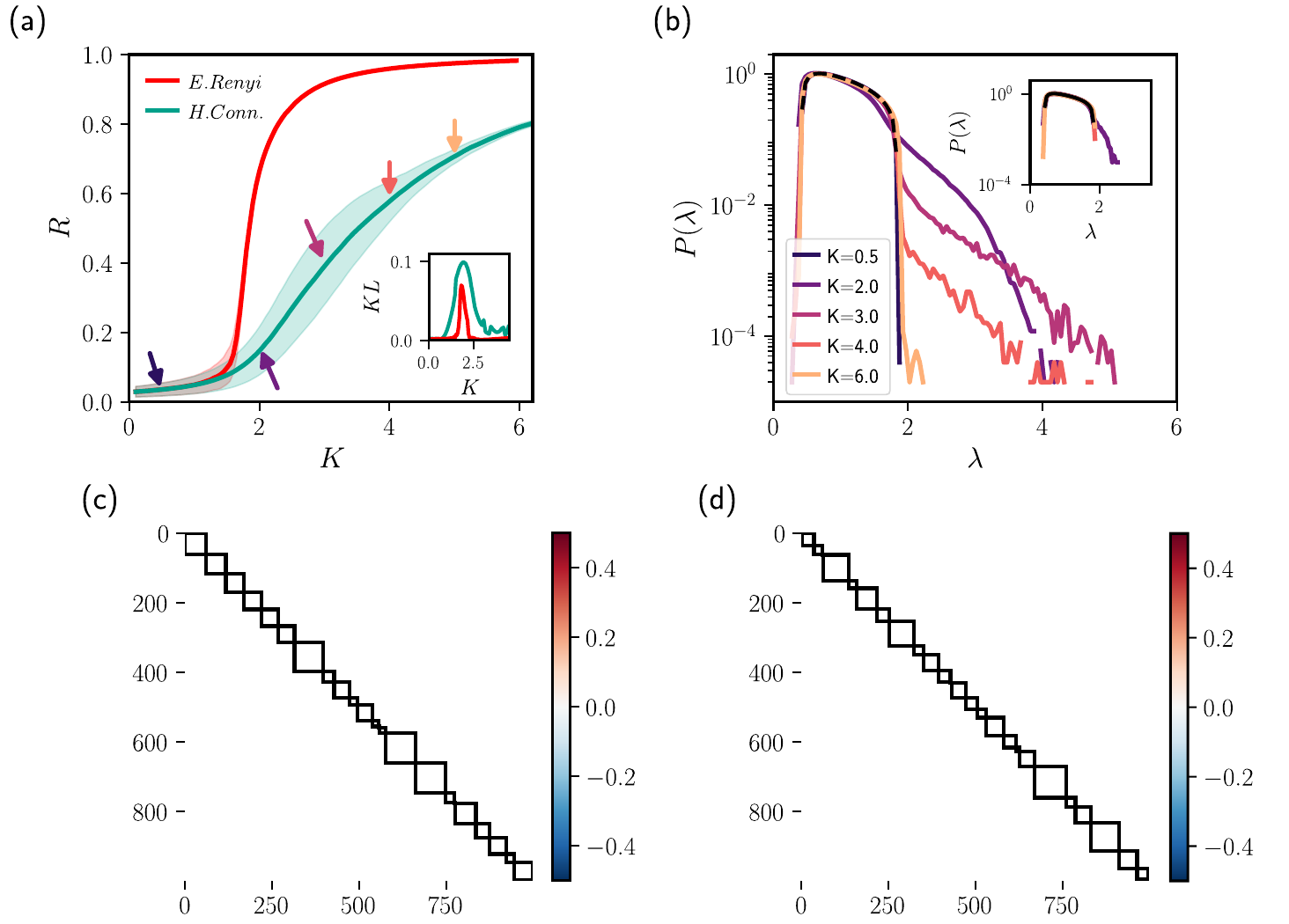}
 \caption{\textbf{Validation of RMT pipeline on a synthetic network model.}
(a) Kuramoto order parameter, $R$, as a function of coupling $K$ for an Erdős-Rényi (ER) network and the human connectome (HC). The HC's heterogeneous structure (used in b-d) creates a complex transition, providing a non-trivial benchmark for emergent correlations.
(b) Eigenvalue distributions for the HC. When weakly coupled ($K=0.1$, inset), the spectrum matches the Marchenko-Pastur (MP) law (dashed line), indicating a successful identification of pure noise. In the coupled regime ($K=3.6$), "signal" eigenvalues (red) clearly emerge, deviating from the MP noise bound.
(c)-(d) Filtered correlation matrices (a common colorbar is used for both matrices). (c) The ER network shows weak structure, while (d) the HC matrix, reconstructed only from the "signal" eigenvalues identified in (b), reveals a strong, non-trivial modular structure. This validates the pipeline's ability to isolate emergent function from noise.}
\label{Kur}
\end{figure}

To complement the non-interacting Random Walk model, we tested our pipeline on a benchmark with known, complex emergent correlations: the Kuramoto model (see SI3). This model simulates synchronization dynamics on a network, governed by:
\[\dot\theta_i = \omega_i + \frac{K}{k_i} \sum_{j=1}^{N} \sin (\theta_j - \theta_i) + \sigma \eta_i(t),
\]
where $K$ is the global coupling strength and $\sigma$ is noise. We simulated this model on a human connectome (Fig.~\ref{Kur}a), which provides a realistic, heterogeneous network structure.

We first validated our derivative-based method. When coupling is zero ($K=0$), the model reduces to independent random walks. As expected, correlations from the raw phases ($\theta_i$) fail, while correlations from the temporal derivatives ($\dot \theta_i$) correctly identify the pure noisy case, matching the Marchenko-Pastur (MP) law (see Fig.~\ref{Kur}b, inset).

Next, we increased the coupling $K$ to a regime where strong, non-trivial correlations emerge from the network's dynamics \cite{Moretti2013,Villegas1,Villegas2} (see also SI3). As shown in Fig.~\ref{Kur}b, the eigenvalue distribution now clearly deviates from the MP-law. These new "signal" eigenvalues, which fall outside the theoretical noise bound, are the signature of true, emergent function.

When we filter out the noise-bound eigenvalues and reconstruct the matrix using only these "signal" components, we recover the network’s underlying modular structure with high fidelity. (Fig.~\ref{Kur}c,d). This validates our RMT pipeline's ability to not only discard noise (as shown in the Random Walk model) but to successfully isolate and preserve true, emergent functional structure (see SI3).

\begin{figure*}[hbtp]
\centering
 \includegraphics[width=2.\columnwidth]{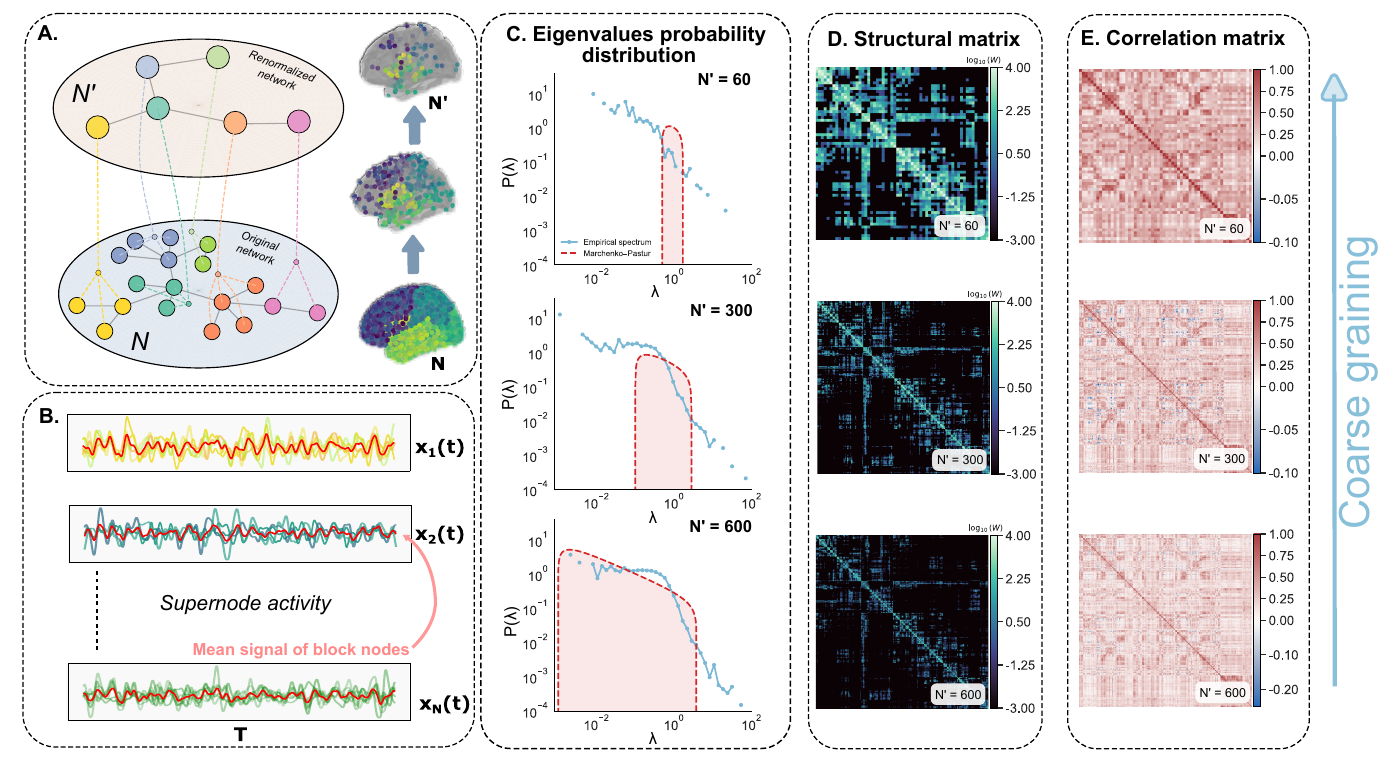}
    \caption{\textbf{Multiscale coarse-graining and spectral filtering of empirical brain networks.}
    \textbf{(A)} Sketch of the Laplacian Renormalization Group (LRG). The LRG is used to hierarchically coarse-grain the human structural connectome from \(N = 2165\) nodes to reduced networks of \(N' = 600\), \(300\), and \(60\) nodes.
    \textbf{(B)} For each reduced network, resting-state fMRI time series (\(T = 652\) points) are averaged within the corresponding supernodes to generate representative signals.
    \textbf{(C)} The eigenvalue probability distributions of the resulting correlation matrices are analyzed. The Marchenko–Pastur law (red curve) defines the theoretical boundary for noise, allowing for the identification and removal of noise-driven eigenvalues.
    \textbf{(D)} Structural connectivity matrices, derived from diffusion MRI, are shown at each coarse-grained resolution.
    \textbf{(E)} Corresponding coarse-grained functional correlation matrices are shown for each resolution, organized by the same structural partitions as in (D).
    Collectively, this workflow integrates structural coarse-graining (A, D) with functional data averaging (B, E) and spectral filtering (C) to produce robust, reduced network models where the number of time points ($T$) is proportional to the network size ($N'$), i.e., $T \sim N'$.}

    \label{fig2}
\end{figure*}

%\textcolor{red}{FALTA PARRAFO DE UNION- BRAIN CRITICO CON REAL DATA}
These benchmarks establish that derivative-based spectral filtering correctly separates noise from emergent correlations, motivating its application to empirical fMRI data, where temporal limitations are unavoidable.

\subsection{Finite time-series in real recordings}
While simulations of the Kuramoto model on the Human Connectome demonstrate how network topology, together with network interacting dynamics, can generate non-trivial correlations, a central challenge in real neural recordings is that they are subject to statistical sampling constraints. Experimental datasets provide a finite number of time points, and practical factors such as participant fatigue, motion artifacts, and acquisition costs limit the achievable sampling depth in EEG and fMRI sessions \cite{Tagliazucchi2014,Power2012,Birn2013}. In real recordings, the sampling regime often violates the assumptions needed for reliable correlation estimation (e.g., the limit \(T \gg N\), where \(T\) is the length of the temporal derivative of the recording and \(N\) is the number of network nodes). Under these circumstances, spurious correlations arising from non-stationarity or finite-size effects can compromise the quality of the measured correlation matrices, making statistical reliability difficult to achieve, as demonstrated by studies showing that scan duration significantly impacts the stability of connectivity estimates \cite{Noble2019, Birn2013}. Applying spectral filtering approaches analogous to those used in theoretical models allows genuine functional interactions to be disentangled from these artifacts. Because $T$ is limited, dimensionality reduction becomes necessary. In practice, this entails coarse-graining the system into a smaller set of representative nodes or modes, so that the effective number of observables satisfies \(T > N_{\mathrm{cg}}\). This reduction enables a reliable estimation of correlations while preserving the essential structure of the underlying neural dynamics.

Our approach is a two-step process to solve the $T < N$ problem. First, we apply LRG to coarse-grain the structural network from $N$ nodes to a much smaller $N'$ nodes. This step ensures that our number of time points $T$ is no longer vastly smaller than our node count, allowing us to achieve a statistically reliable $T \sim N'$ regime. Second, having created this reliable $N' \times N'$ system, we then apply Random Matrix Theory (RMT), using the Marchenko-Pastur law to identify and filter the noise-dominated eigenvalues from this new, coarse-grained correlation matrix. This leaves only the true, non-noise components of the functional connectivity.

% Figure \ref{XX} reports the example of... \textcolor{red}{Fig.3 Temporal series. Functional correlation matrix. Eigenvalues. WTF is the Marchenko-Pastur here?  Problem of time resolution $T<N$.}

\subsubsection{Coarse-graining limited temporal data}

We reduce dimensionality by coarse-graining nodes using structural information only. Importantly, partitions must be defined independently of functional correlations. This is because functional correlations are the emergent properties we aim to measure; using them to define the coarse-graining partitions would be a circular argument that undermines the analysis.

Therefore, drawing inspiration from multiscale methods in statistical physics \cite{Kadanoff66}, we group nodes based on their structural relationships. This strategy controls how activity flows across scales and, most importantly, preserves the network’s true multiscale organization. The Laplacian Renormalization Group (LRG) \cite{LRG,jstat_statphys} is the formal framework that implements this strategy, providing a principled, data-driven approach to define supernodes enabling a robust estimation of coarse-grained functional interactions.

The LRG procedure can be summarized in two main steps (see Methods and SI4 for a detailed description). First, the network is represented as a structural graph, with nodes corresponding to structural brain regions and edges encoding anatomical connectivity. Starting from the adjacency matrix $\hat A$ and the diagonal degree matrix $\hat D$, we compute the graph Laplacian $\hat L = \hat D - \hat A$ and the associated diffusion propagator $\hat K(\tau) = e^{-\tau \hat L}$. This defines the \emph{communicability matrix}. Using its trace-normalized form $\rho(\tau)=e^{-\tau \hat L}/\mathrm{Tr}\,e^{-\tau \hat L}$, we define the \emph{communication distance} as $\mathcal{D}_{ij}(\tau)=(1-\delta_{ij})/\rho_{ij}(\tau)$ \cite{Modularity,LRG,InfoCore}. Second, for a fixed $\tau$, these distances define a hierarchical dendrogram of the network, enabling a controlled multiscale coarse-graining in which structurally consistent groups of nodes are merged step by step \cite{Modularity}. Here, $\tau$ acts as a continuous scale parameter: small values preserve fine structural detail by emphasizing microscopic Laplacian modes, while larger values progressively attenuate these modes, revealing broader, more global communities. To retain maximal network detail, we focus on small $\tau$ and cut the dendrogram at the level where the reduced network contains $N'$ supernodes, ensuring $T > N'$.

Finally, functional correlations are estimated at the coarse-grained (``Kadanoff'') supernodes level, providing a statistically reliable measure of interactions that reflects the underlying structural constraints. Within each supernode, the activity is averaged as $\bar X_i = \frac{1}{N_b}\sum_{j=1}^{N_b} X_j,$ where \(N_b\) is the number of constituent nodes within the supernode. This procedure reduces a heterogeneous network of \(N\) neural populations to one with \(N'\) supernodes, ensuring that the statistical condition \(T > N'\) is satisfied while preserving the essential structure of the network.

We applied this LRG-RMT procedure to an empirical human connectome (see Methods), with a structural network of \(N = 2165\) nodes and fMRI time series of \(T = 652\) time points. The original $T \ll N$ regime makes the correlation matrix statistically unreliable and dominated by noise.

Our two-step process addresses this. First, we applied the LRG (Fig.~\ref{fig2}A) to coarse-grain the structural network from $N=2165$ to different final resolutions of $N' = 600$, $300$, and $60$ nodes (see also Fig.~\ref{fig2}D). The fMRI time series were then averaged within these supernodes (see Fig.~\ref{fig2}B) to create new functional matrices (Fig.~\ref{fig2}E) that satisfy the $T \sim N'$ condition. Second, we applied RMT to these new, statistically reliable matrices. We plotted their eigenvalue distributions (Fig.~\ref{fig2}C, empirical curve) against the theoretical Marchenko-Pastur (MP) law (Fig.~\ref{fig2}C, red curve), which defines the boundary for a network composed purely of noise. This analysis revealed two key results. First, the empirical distributions clearly deviate from the MP-law; the eigenvalues falling outside the theoretical noise bound represent the true, non-noise functional correlations. Second, as the network dimension is reduced from $N'=600$ to $N'=60$, the noise bound width of the Marchenko-Pastur law (the $\lambda_+$ value) becomes progressively smaller, demonstrating a significant reduction in the noise-dominated portion of the signal.

Crucially, the coarse-grained functional connectivity matrices (without spectral filtering, see Fig.~\ref{fig2}E) preserve --at all scales-- the large-scale topological structure of the original network. This shows that LRG preserves multiscale organization across resolutions, while subsequent spectral filtering removes noise-dominated components.

%The structural adjacency matrix from diffusion MRI defines the graph Laplacian \(\hat L\), which governs the diffusion propagator and the resulting LRG hierarchy.
%At each renormalization step, structurally defined nodes were merged into supernodes according to the communication distance derived from the propagator, and the BOLD signals within each supernode were averaged to produce representative functional time series. The dendrogram was cut to ensure that the reduced network satisfied the statistical condition \(T > N'\), yielding successive resolutions of \(N' = 600\), \(N' = 300\), and \(N' = 60\) (see Figure~\ref{fig2}). \textcolor{red}{improve and explain figure in detail-->}Importantly, the eigenvalue distributions of the correlation matrices of the temporal derivatives deviate from the Marchenko–Pastur prediction, indicating that correlations cannot be explained purely by noise. As the dimensionality is reduced, the range of eigenvalues compatible with random fluctuations shrinks, consistent with the constraint imposed by Eq.~\ref{MPastur}.

\begin{figure*}[htpb]
\centering
\includegraphics[width=2\columnwidth]{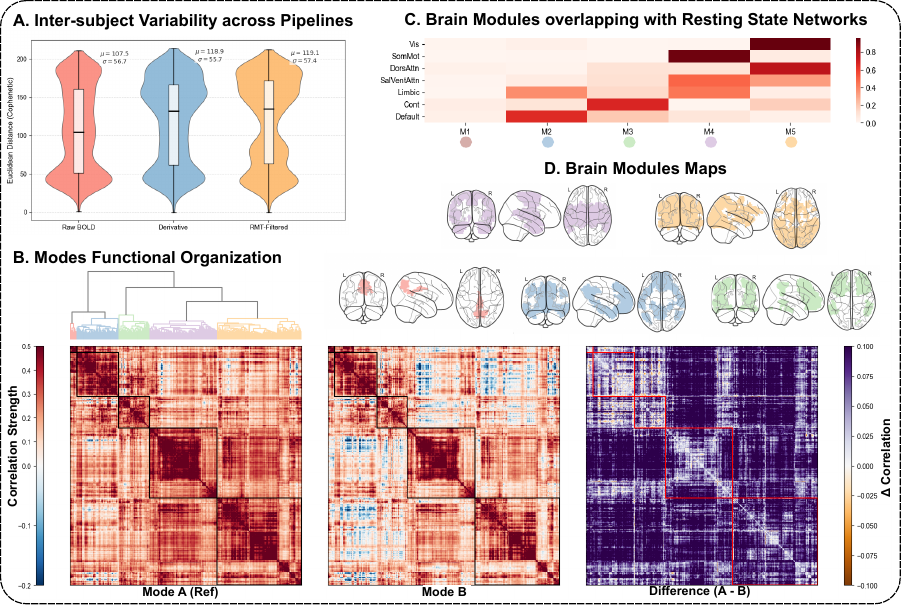}
\caption{\textbf{LRG--RMT pipeline reveals hidden functional organizational modes and their neuroanatomical signatures.}
\textbf{(A)} Distributions of pairwise inter-subject distances computed from LRG-derived networks at an intermediate resolution ($N' \approx 300$). Functional connectivity (FC) estimated from raw BOLD signals (1) yields a narrow, unimodal distribution ($\mu = 107.5$), indicating limited apparent inter-subject variability. Using temporal derivatives (2) shifts the distribution toward higher mean distances but remains weakly structured. In contrast, the LRG--RMT pipeline (3) produces a broad, multimodal distribution ($\mu = 119.1$), revealing previously obscured population structure.
\textbf{(B)} Hierarchical clustering of the RMT-filtered FC matrices identifies two dominant functional organizational modes (Mode A and Mode B). While both modes share the same underlying structural modularity (modules M1--M5), they differ in their patterns of inter-modular functional organization. The difference matrix ($A - B$) highlights pronounced inter-modular anticorrelations in Mode B (red boxes), which are attenuated in Mode A.
\textbf{(C)} Spatial correspondence between LRG-derived structural modules and canonical resting-state networks (RSNs), demonstrating alignment with established functional systems (e.g., Visual, Default Mode, and Salience networks). 
\textbf{(D)} Neuroanatomical projection of the five structural modules (M1--M5) identified by LRG. Mode A: $n=96$; Mode B: $n=40$.}
\label{fig:biotypes}
\end{figure*}

\subsection{LRG-RMT uncovers hidden functional organization modes}
Having established the LRG-RMT procedure, we next examined how methodological choices in functional connectivity (FC) estimation shape population-level structure after coarse-graining. Specifically, we asked whether different FC pipelines lead to qualitatively different representations of inter-subject variability once network dimensionality is reduced to a statistically reliable scale.

We analyzed the same empirical dataset ($N = 2165$ nodes, $T = 652$ time points) for all subjects, applying LRG to coarse-grain the structural network to an effective resolution of $N' \sim 300$. For each subject, we constructed three FC matrices: (1) correlations of raw BOLD signals, (2) correlations of temporal derivatives, and (3) derivative-based correlations filtered using Random Matrix Theory (RMT). Population structure was quantified using pairwise inter-subject distances between dendrograms derived from the positive backbone of each FC matrix (see Methods).

As shown in Figure~\ref{fig:biotypes}A, FC estimation choices profoundly alter the inferred population landscape. Raw BOLD correlations yield a narrow, unimodal distribution of inter-subject distances ($\mu = 107.5$, $\sigma = 56.7$), consistent with FC matrices dominated by global, non-stationary fluctuations that artificially homogenize subjects. Using temporal derivatives partially mitigates this effect, shifting the distribution toward higher mean distances ($\mu = 118.9$) but still failing to reveal clear population structure. In contrast, combining LRG coarse-graining with RMT filtering produces a qualitatively different outcome: the distribution becomes broad and distinctly multimodal ($\mu = 119.1$, $\sigma = 57.4$), revealing previously hidden structure within the population.

We refer to these patterns as functional organization modes, denoting distinct large-scale organizations of inter-regional interactions at the level of time-averaged connectivity that emerge once noise-dominated components are removed. We emphasize that these modes are presented as an existence proof of hidden functional diversity revealed by the pipeline, rather than as a definitive classification of the healthy population.

Hierarchical clustering of the RMT-filtered FC matrices consistently identifies two dominant regimes (labeled A and B in Figure~\ref{fig:biotypes}B). To assess their anatomical grounding, we mapped the underlying structural modules (M1–M5) to canonical resting-state networks (Figure~\ref{fig:biotypes}C–D), revealing a strong correspondence between the data-driven modules and established systems, including visual, somatomotor, and default-mode networks.

While both regimes share similar anatomical modularity, they differ markedly in their patterns of inter-modular functional organization. As highlighted by the difference matrix ($A - B$), Mode B exhibits structured anticorrelations between specific modules—most notably between the Default Mode Network and task-positive systems—consistent with a segregated functional organization \cite{Fox2005}. In contrast, Mode A shows a pronounced attenuation or absence of such anticorrelations, indicative of a more globally integrated or hypersynchronized organizational pattern.

Together, these results demonstrate that rigorous noise filtering, enabled by structural coarse-graining, can qualitatively reshape the population structure inferred from functional connectivity, uncovering distinct large-scale organizational regimes that remain obscured under standard analysis pipelines.

%Our LRG-RMT pipeline revealed two distinct functional biotypes that differ primarily in their emergent functionality. While Biotype A is characterized by strong positive correlations (integration) within modules, Biotype B exhibits a widespread pattern of negative correlations (anticorrelations).

%In resting-state dynamics, such negative correlations are a signature of functional segregation \cite{Fox2005, Uddin2009}. They typically reflect competitive or orthogonal relationships between large-scale networks, most notably the 'push-pull' dynamic between the Default Mode Network (DMN) and task-positive systems. The prevalence of these anticorrelations in Biotype B suggests a neural architecture with sharper modular boundaries and stronger mutual inhibition between functional communities.

%Crucially, standard FC pipelines often obscure these negative weights by thresholding or global signal regression artifacts. By preserving these features, our method reveals that a significant portion of the population (Biotype B) operates in a highly segregated dynamical regime, distinct from the more integrated regime of Biotype A.

\section{DISCUSSION AND CONCLUSIONS}

Estimating reliable functional connectivity from temporally limited neuroimaging data remains a central challenge in systems neuroscience. By combining Laplacian Renormalization Group (LRG) coarse-graining with Random Matrix Theory (RMT) spectral filtering, we introduce a framework that satisfies the statistical requirements for covariance estimation ($T > N'$) while preserving the multiscale organization of brain networks. This integrated approach reveals population structure that remains inaccessible to standard pipelines, which tend to collapse inter-subject variability into an artificially homogeneous representation.

The functional organization modes uncovered by the LRG-RMT pipeline differ primarily in their patterns of large-scale integration rather than in their underlying anatomical modularity. While both modes share a common structural organization, they exhibit markedly different inter-modular organization. In particular, one mode displays pronounced anticorrelations between the Default Mode Network and task-positive systems, a hallmark of functional segregation and push--pull dynamics associated with efficient information processing \cite{diez_novel_2015}. Crucially, our LRG framework does not rely on negative correlations driven by Global Signal Regression (GSR); instead, it combines diffusion-informed structural coarse-graining (LRG) with spectral noise filtering (RMT). This allows us to recover robust inter-subject variability in both positively and negatively correlated networks without explicitly removing a global mean signal. Taken together, these results indicate that the anticorrelations previously debated \cite{Fox2005,Murphy2009,Murphy2017} can emerge as robust, subject-specific features of the functional connectome, rather than arising solely as mathematical or preprocessing artifacts, consistent with recent evidence for a structural basis of antagonistic networks \cite{Martinez2022}. In contrast, the other mode is characterized by an attenuation or absence of these anticorrelations, indicative of a more globally integrated or hypersynchronized organizational pattern. These differences are largely obscured by standard correlation-based analyses.

Importantly, our results highlight the scale-dependent nature of functional differentiation across subjects. The signatures distinguishing these functional regimes emerge most clearly at intermediate, mesoscopic resolutions ($N' \approx 300$), while they progressively vanish both near the statistical resolution limit ($N' \approx 600$) and at coarser scales ($N' \approx 60$), where subjects converge toward a common functional backbone (see SI5 and Supplementary Figs. S4–S8). This observation suggests that while the macroscopic architecture of the human brain is broadly conserved, meaningful inter-subject variability is encoded in specific mesoscale circuit interactions. The LRG framework is uniquely suited to identify this optimal resolution, as it groups regions according to intrinsic diffusion geometry rather than arbitrary spatial proximity. Our findings align with recent macroscale theories suggesting that functional modes are fundamentally shaped by the underlying structural eigenmodes \cite{Fotiadis2024}. However, unlike static mapping approaches \cite{Turnbull2025, Fotiadis2025}, our renormalization scheme dynamically utilizes this structure–function interplay to filter noise. In this sense, our framework operationalizes the multiscale perspective advocated in network neuroscience, enabling data-driven transitions across organizational scales rather than fixing a priori resolutions \cite{Bassett2017}.

While the present results demonstrate that standard pipelines can suppress genuine inter-subject variability, the prevalence, temporal stability, and behavioral relevance of the uncovered functional organizational modes cannot be established from the current dataset alone. Our sample size ($N = 136$) is sufficient to demonstrate the methodological effect, but not to estimate population frequencies or to determine whether these modes reflect stable traits or transient functional states. Addressing these questions will require larger cohorts and longitudinal measurements.

In summary, functional connectivity estimation choices act as a lens that shapes the population structure that becomes observable. By rigorously addressing the $T < N$ limitation and filtering noise while preserving network architecture, the LRG-RMT framework provides a principled foundation for future studies aimed at linking large-scale dynamical modes to cognition, behavior, and pathology.

\section{METHODS}

\paragraph*{\textbf{Empirical datasets and preprocessing.}}Two human connectome datasets were used for complementary purposes in this study. For the Kuramoto model simulations shown in Fig.~\ref{Kur}, we used a previously published human structural connectome \cite{Sporns2005}, comprising $N = 998$ nodes with average degree $\langle k \rangle \approx 30$.

The analysis of inter-subject variability using the LRG--RMT pipeline (Figs.~\ref{fig2} and \ref{fig:biotypes}) was performed on a high-resolution multimodal dataset from \cite{JimenezMarin2024}. This dataset includes diffusion MRI-derived structural connectivity and resting-state fMRI time series for 136 healthy subjects. Structural networks were parcellated at the highest available resolution ($N = 2165$ nodes). Resting-state fMRI time series were preprocessed following the protocols detailed in \cite{JimenezMarin2024}, which utilized the preprocessed rs-fMRI acquisitions from the MPI-LEMON dataset \cite{Babayan2018}. These pipelines included motion correction, co-registration, normalization to MNI space, and nuisance regression, resulting in time series of length $T = 652$. Structural and functional preprocessing were performed independently.

\paragraph*{\textbf{Laplacian Renormalization Group coarse-graining.}}
Network dimensionality was reduced using the Laplacian Renormalization Group (LRG), a diffusion-based coarse-graining framework for complex networks \cite{LRG,jstat_statphys}. For each subject, the graph Laplacian $\hat L = \hat D - \hat A$ was constructed from the structural adjacency matrix $\hat A$ and degree matrix $\hat D$. Diffusion on the network follows the standard heat-kernel dynamics on graphs $\dot{\mathbf{s}}(\tau) = -\hat L \mathbf{s}(\tau)$, with propagator $\hat K(\tau) = e^{-\tau \hat L}$. 

In practice, distances are computed from the trace-normalized propagator\cite{de_domenico_spectral_2016,InfoCore,Modularity}, $\rho_{ij}(\tau)=K_{ij}(\tau)/\mathrm{Tr}\,K(\tau)$, so that communication distances are defined up to a global $\tau$-dependent scale factor without affecting the hierarchical clustering \cite{Modularity,InfoCore}.

LRG groups nodes into supernodes based on similarities in diffusion geometry, producing a multiscale, data-driven parcellation without relying on predefined anatomical atlases \cite{LRG,jstat_statphys}. Coarse-grained networks of size $N'$ were obtained by hierarchical clustering of diffusion-based node distances \cite{Modularity,InfoCore}, yielding effective resolutions of $N' = 600$, $300$, and $60$. Unless otherwise stated, results in the main text focus on the intermediate scale $N' \approx 300$, for which covariance estimation becomes statistically well-conditioned given the available time series length ($T > N'$).

Dimensionality reduction is ubiquitous in fMRI analyses, including PCA-, ICA-, and gradient-based manifold approaches \cite{Beckmann2004, Margulies2016}. However, these methods typically operate on functional signals alone. In contrast, LRG defines reductions using structural diffusion geometry, ensuring that coarse-graining reflects anatomical communication constraints rather than statistical variance alone. This structurally grounded reduction is critical when the goal is statistically reliable covariance estimation rather than component separation.

\paragraph*{\textbf{Functional connectivity estimation.}} Functional connectivity (FC) was computed on the LRG-reduced networks using three pipelines. 
First, FC matrices were estimated using Pearson correlations between raw BOLD time series. 
Second, FC matrices were computed using correlations of temporal derivatives of the BOLD signals, which reduces the influence of slow global fluctuations. 
Third, derivative-based FC matrices were further processed using Random Matrix Theory (RMT) spectral filtering to remove noise-dominated components.

All FC matrices were symmetrized before further analysis. No thresholding was applied unless explicitly stated.

\paragraph*{\textbf{Random Matrix Theory spectral filtering.}} Noise filtering was performed using Random Matrix Theory (RMT) applied to the derivative-based FC matrices. The empirical eigenvalue spectrum of each correlation matrix was compared to the Marchenko--Pastur distribution expected for random covariance matrices with identical dimensions and sampling ratios. Eigenmodes whose eigenvalues lay within the RMT-predicted noise bulk were identified as noise-dominated and removed, while outlying modes were retained to reconstruct a filtered FC matrix. This procedure yields a statistically controlled estimate of functional connectivity that explicitly accounts for finite time-series effects.

\paragraph*{\textbf{Population-level analysis.}} Population structure was quantified using pairwise inter-subject distances between dendrograms derived from the positive backbone of FC matrices. Specifically, for each subject, negative FC entries were set to zero only for the purpose of constructing a stable similarity metric for hierarchical clustering. Inter-subject dissimilarity was then computed as the euclidean distance between the cophenetic distance matrices associated with each dendrogram.

Importantly, this representation is used solely to define distances between subjects. All reported functional organization modes, difference matrices, and neurobiological interpretations are computed from the full signed FC matrices, preserving anticorrelations.

\paragraph*{\textbf{Neuroanatomical mapping.}} Structural modules identified by LRG were projected back onto cortical and subcortical anatomy for visualization. To assess correspondence with established functional systems, LRG-derived modules were compared with canonical resting-state networks using spatial overlap measures. This mapping was used solely for anatomical interpretation and did not influence the identification of functional organizational modes.

\paragraph*{\textbf{Data Availability.}} All structural and functional neuroimaging data used in the inter-subject variability analysis are openly available in the repository described by Jimenez-Marin \emph{et al.}, \emph{Open datasets and code for multi-scale relations on structure, function, and neuro-genetics in the human brain}, \textit{Scientific Data} 11, 256 (2024), DOI: \href{https://doi.org/10.1038/s41597-024-03060-2}{10.1038/s41597-024-03060-2}. The structural connectome used for the Kuramoto simulations is available from the original publication by Sporns \emph{et al.} \cite{Sporns2005}.

\vspace{-0.5cm}\subsection*{Acknowledgments}\vspace{-0.2cm}
I.F.I. acknowledges financial support from a predoctoral grant from the Basque Government (PRE\_2025\_2\_0208). P.V. acknowledges the Spanish Ministry of Research and Innovation and Agencia Estatal de Investigaci\'on (AEI), MICIN/AEI/10.13039/501100011033, for financial support through Project PID2023-149174NB-I00, funded also by European Regional Development Funds, as well as Ref. PID2020-113681GB-I00.  JMC acknowledges financial support from Ikerbasque: The Basque Foundation for Science, and from Spanish Ministry of Science (PID2023-148008OB-I00), Spanish Ministry of Health (PI22/01118), Basque Ministry of Health (grants 2022111031,  2023111002  \& 2025111091).

\section*{Competing interests}
The authors declare no competing interests.

\section*{Author contributions}
J.C., P.V., and I.F.-I conceived the study. J.C. and P.V. supervised the study. I.F.-I. performed the analyses. A.J.-M. provided the datasets and pre-processing resources. P.V. developed the theoretical framework and analytical tools. I.F.-I., J.C., and P.V. interpreted the results. P.V. and I.F.-I. wrote the manuscript. All authors discussed the results, reviewed, and approved the final manuscript.

\def\url#1{}
%\bibliographystyle{plain}
%\bibliography{CorrMat} 

\clearpage


\section*{Supplementary material}

# Supplementary Information: Structural coarse-graining enables noise-robust functional connectivity and reveals hidden inter-subject variability

Izaro Fernandez-Iriondo,[1, 2, 3, *] Antonio Jimenez-Marin,[2] Jesus Cortes,[2, 4, 5] and Pablo Villegas[6, 7, †]

[1]*Computer Science and Artificial Intelligence, University of the Basque Country (UPV/EHU), San Sebastian, Spain*
[2]*Computational Neuroimaging Lab, Biobizkaia Health Research Institute, Barakaldo, Spain*
[3]*Informatics Engineering Doctoral Programme, University of the Basque Country (UPV/EHU), San Sebastian, Spain*
[4]*IKERBASQUE: The Basque Foundation for Science, Bilbao, Spain*
[5]*Department of Cell Biology and Histology, University of the Basque Country (UPV/EHU), Leioa, Spain*
[6]*'Enrico Fermi' Research Center (CREF), Via Panisperna 89A, 00184 - Rome, Italy*
[7]*Instituto Carlos I de Física Teórica y Computacional, Univ. de Granada, E-18071, Granada, Spain.*

## CONTENTS

---

* fernandeziriondo.izaro@gmail.com
† pablo.villegas@cref.it

## 1. SPECTRAL PROPERTIES OF SPURIOUS CORRELATIONS IN RANDOM WALKS

This section provides intuition for why non-stationary signals can generate a singular eigenvalue density near the origin even in the absence of true interactions. We illustrate how random walks produce spurious correlations and show that this behavior follows from the spectral properties of the underlying integration operator.

Let $x_i(t)$ be $N$ independent $T$-step random walks defined as cumulative sums of i.i.d. increments $\xi_i(s)$ with zero mean and unit variance:

$$x_i(t) = \sum_{s=1}^{t} \xi_i(s).$$

Writing the data matrix $Y \in \mathbb{R}^{N \times T}$ as $Y_{it} = x_i(t)$, we factorize

$$Y = XL^\top,$$

where $X_{is} = \xi_i(s)$ and $L$ is the discrete integration operator ($L_{ts} = 1$ if $s \leq t$, otherwise 0). The sample covariance reads

$$C = \frac{1}{T} YY^\top = \frac{1}{T} X \mathcal{T} X^\top, \qquad \mathcal{T} = L^\top L.$$

While $\mathcal{T}$ differs from the direct covariance of the walk ($LL^\top$), the two are isospectral and therefore share the same large-$T$ spectral scaling. After normalization $\tilde{\mathcal{T}} = \frac{1}{T^2}\mathcal{T}$, its spectrum converges to that of the Brownian covariance operator on $[0,1]$ with kernel $K(u,v) = \min(u,v)$. The eigenvalues obey

$$\tilde{\tau}_k \approx \frac{1}{\pi^2(k-1/2)^2} \sim k^{-2}.$$

Treating $k$ as continuous gives the density of eigenvalues of the temporal covariance operator

$$\rho_{\tilde{\mathcal{T}}}(\tilde{\tau}) \propto \tilde{\tau}^{-3/2},$$

valid for $T^{-2} \ll \tilde{\tau} \ll O(1)$.

Thus, temporal integration generates a strongly inhomogeneous spectrum with a slow $k^{-2}$ decay, leading to a heavy accumulation of small eigenvalues. In correlated Wishart ensembles, the sample spectrum is controlled by the population spectrum [1, 2]. Consequently, this heavy accumulation of small temporal modes can induce a $\lambda^{-3/2}$-type regime in the empirical covariance spectrum, consistent with our numerical results. Pearson normalization (rescaling rows and columns by inverse standard deviations) does not change this scaling at leading order in the large-$N$,$T$ limit, since the spectral singularity arises from the temporal operator $\mathcal{T}$ common to all walkers.

Finally, differencing the series,

$$z_i(t) = x_i(t) - x_i(t-1),$$

effectively removes the integration operator ($\mathcal{T} \to I$), yielding

$$C \approx \frac{1}{T} XX^\top,$$

which follows the standard Marchenko–Pastur law for stationary increments [2, 3].

## 2. NON-INTERACTING CONFINED RANDOM WALKS

To test whether temporal differentiation is required only for non-stationary signals, we analyzed the Ornstein–Uhlenbeck (OU) process, the canonical stationary and mean-reverting stochastic process:

$$\dot{x}_i(t) = -ax_i(t) + \sigma\eta_i(t).$$

We simulated $N = 1000$ non-interacting OU processes and computed eigenvalue spectra of their correlation matrices. Despite stationarity, raw correlation matrices deviate strongly from the Marchenko–Pastur prediction due to temporal autocorrelations (Supplementary Fig. 1a). In contrast, correlation matrices computed from temporal derivatives recover the MP distribution accurately (Supplementary Fig. 1b).

This demonstrates that temporal differentiation acts as an effective whitening step for both stationary and non-stationary autocorrelated signals, ensuring that RMT filtering removes only noise.

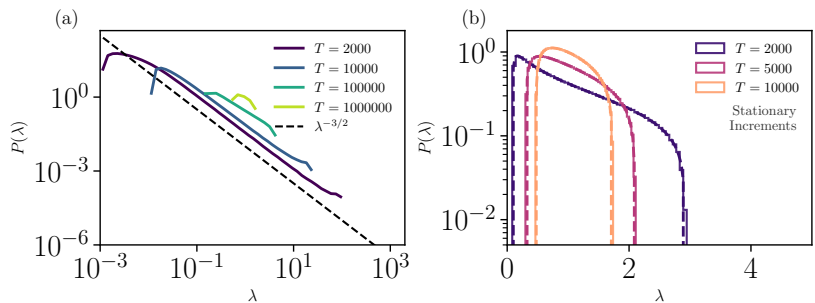

FIG. 1. **Analysis of stationary (confined) signals. (a)** Eigenvalue probability distribution for the Pearson correlation matrix of $N = 1000$ non-interacting Ornstein–Uhlenbeck processes. Despite stationarity, temporal autocorrelations generate spurious eigenvalues deviating from the Marchenko–Pastur (MP) bound. **(b)** Eigenvalue distribution for the temporal derivatives of the same signals, which accurately matches the MP distribution. *Parameters: $a = \sigma = 1$, $dt = 10^{-2}$. All curves averaged over $10^2$ realizations.*

## 3. CRITICAL DYNAMICS, KURAMOTO MODEL, AND GRIFFITHS PHASES

To illustrate how genuine functional correlations can emerge from interacting dynamics on networks, we simulated a Kuramoto model on the human connectome:

$$\dot{\theta}_i = \omega_i + \frac{K}{k_i}\sum_{j=1}^{N}\sin(\theta_j - \theta_i) + \sigma\eta_i(t),$$

where $\theta_i$ is the phase of node $i$, $\omega_i$ the intrinsic frequency, $K$ the coupling strength, and $\eta_i(t)$ Gaussian noise.

The Kuramoto model provides a convenient testbed because it generates correlations that arise purely from interactions on a known structural network. As coupling increases, the system transitions from largely independent fluctuations to collective dynamics that produce structured correlation patterns. Near the synchronization transition, fluctuations become long-ranged and correlations extend across the network.

Importantly, in heterogeneous networks such as the human connectome, structural disorder broadens the transition region, giving rise to Griffiths-like regimes characterized by frustrated or partial synchronization [4–6]. In these regimes, correlations can be spatially structured yet temporally intermittent, creating a challenging scenario for distinguishing genuine interactions from noise.

Here, criticality is not invoked as a hypothesis about brain dynamics, but as a controlled regime in which interaction-driven correlations are known to exist. This makes it a stringent benchmark for testing whether the proposed pipeline can correctly separate interaction-induced structure from sampling noise.

In the limiting case $K = 0$, the dynamics reduce to independent random walks. In this regime, Pearson correlations computed from $\theta_i(t)$ produce spurious correlations, whereas correlations computed from $\dot{\theta}_i(t)$ converge to the Marchenko–Pastur prediction, confirming that the pipeline correctly identifies the absence of true interactions.

## 4. DETAILED FORMULATION OF THE LAPLACIAN RENORMALIZATION GROUP

The Laplacian Renormalization Group (LRG) coarse-graining procedure is based on network diffusion governed by the graph Laplacian $\hat{L} = \hat{D} - \hat{A}$. Diffusion dynamics follow $\dot{\mathbf{s}}(\tau) = -\hat{L}\mathbf{s}(\tau)$, whose solution defines the network propagator $\hat{K}(\tau) = e^{-\tau \hat{L}}$.

From this propagator, we define the Laplacian density matrix

$$\hat{\rho}(\tau) = \frac{e^{-\tau \hat{L}}}{\operatorname{Tr} e^{-\tau \hat{L}}},$$

which provides a trace-normalized representation of diffusion-based communication across the network.

Node-to-node distances are then defined as

$$\mathcal{D}_{ij}(\tau) = \frac{1 - \delta_{ij}}{\rho_{ij}(\tau)},$$

so that nodes with stronger diffusion-mediated communication are considered closer. These distances define a hierarchical geometry on the network.

Hierarchical clustering using Ward's linkage was applied to this distance matrix to construct dendrograms, which were cut at different levels to produce coarse-grained networks of size $N'$.

## 5. SIGNALS OF VARIABILITY ACROSS SCALES

In this section, we examine how inter-subject functional variability depends on the observation scale imposed by structural coarse-graining. Because LRG explicitly defines a multiscale representation of the connectome, it is important to assess how the inferred population structure depends on the chosen resolution. While the main text focuses on the intermediate resolution ($N' \approx 300$), where reliable covariance estimation and maximal population differentiation coexist, here we evaluate the robustness of these findings across nearby scales.

To this end, we analyze functional connectivity patterns obtained at alternative coarse-graining resolutions and compare the effects of standard (raw BOLD) and RMT-filtered pipelines. Our goal is not to re-identify functional organizational modes at each scale, but rather to assess the robustness and scale-dependence of the structured functional patterns associated with Mode B.

At finer resolutions, increased dimensionality amplifies finite-sample noise, whereas at coarser resolutions structural averaging progressively suppresses inter-subject differences. Scale, therefore, plays a critical role in shaping the balance between noise suppression and sensitivity to functional heterogeneity. This trend is consistent with the expected scaling of sampling noise with dimensionality ($\sim N'/T$).

Together, these analyses show that maximal inter-subject differentiation emerges at intermediate resolutions, justifying the focus on $N' \approx 300$ in the main text.

### A. Comparison of Mode B functional connectivity pattern at the $N' = 391$ scale

Figure 2 compares the functional connectivity pattern associated with Mode B at the resolution $N' = 391$, computed using raw BOLD correlations (Pipeline 1) and the RMT-filtered pipeline (Pipeline 3), together with their difference. While the raw BOLD pipeline yields a weakly structured and spatially diffuse pattern, the RMT-filtered pipeline reveals a clear and well-defined functional organization. The difference matrix highlights structured anticorrelations that become apparent after noise filtering, demonstrating that the Mode B organization pattern persists at nearby intermediate resolutions.

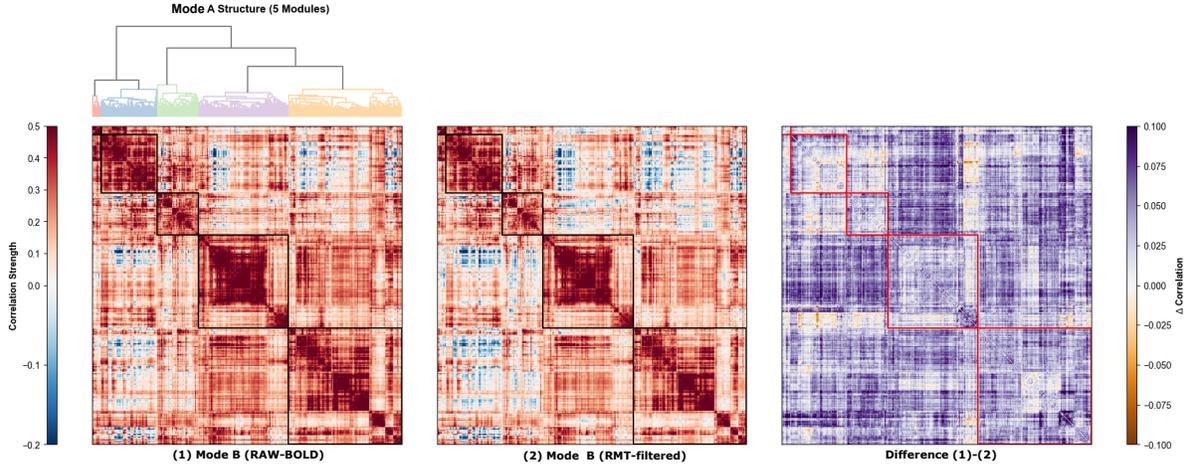

FIG. 2. **Robustness of Mode B functional organization at an intermediate scale ($N' = 391$).** **Left:** Functional connectivity matrix computed from raw BOLD correlations (Pipeline 1). **Center:** Functional connectivity matrix computed using the LRG–RMT pipeline (Pipeline 3). **Right:** Difference between the two matrices, highlighting structured functional patterns revealed by noise filtering.

*Brain modules description based on anatomical atlas*

To assess the anatomical interpretability of the LRG-derived structural modules, we quantified their overlap with regions defined by the Desikan–Killiany anatomical atlas. Figure 3 shows the correspondence between data-driven modules (M1–M5) and anatomical regions, indicating that each module spans multiple contiguous cortical areas rather than reflecting a single anatomical parcel.

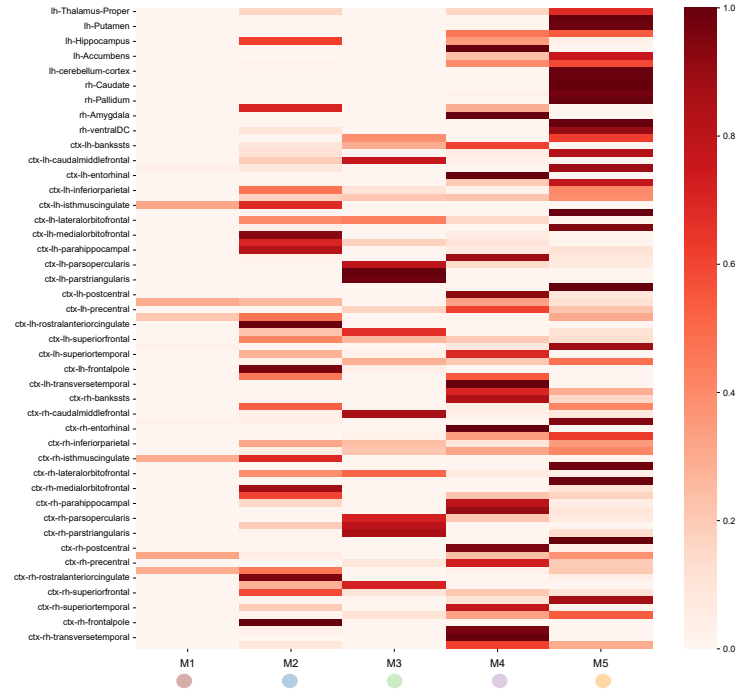

FIG. 3. **Anatomical characterization of LRG-derived modules.** Heatmap showing the overlap between data-driven structural modules (M1–M5, columns) and anatomical regions defined by the Desikan–Killiany atlas (rows).

## B.  LRG–RMT analysis at the $N' = 600$ scale

We next repeated the LRG–RMT analysis at a finer resolution ($N' = 600$), where network dimensionality approaches the temporal sampling limit. Figure 4 illustrates the population-level structure revealed at this scale. While RMT filtering continues to enhance inter-subject variability relative to standard pipelines, population structure appears less sharply differentiated than at intermediate resolutions, reflecting increased sensitivity to finite-sample noise.

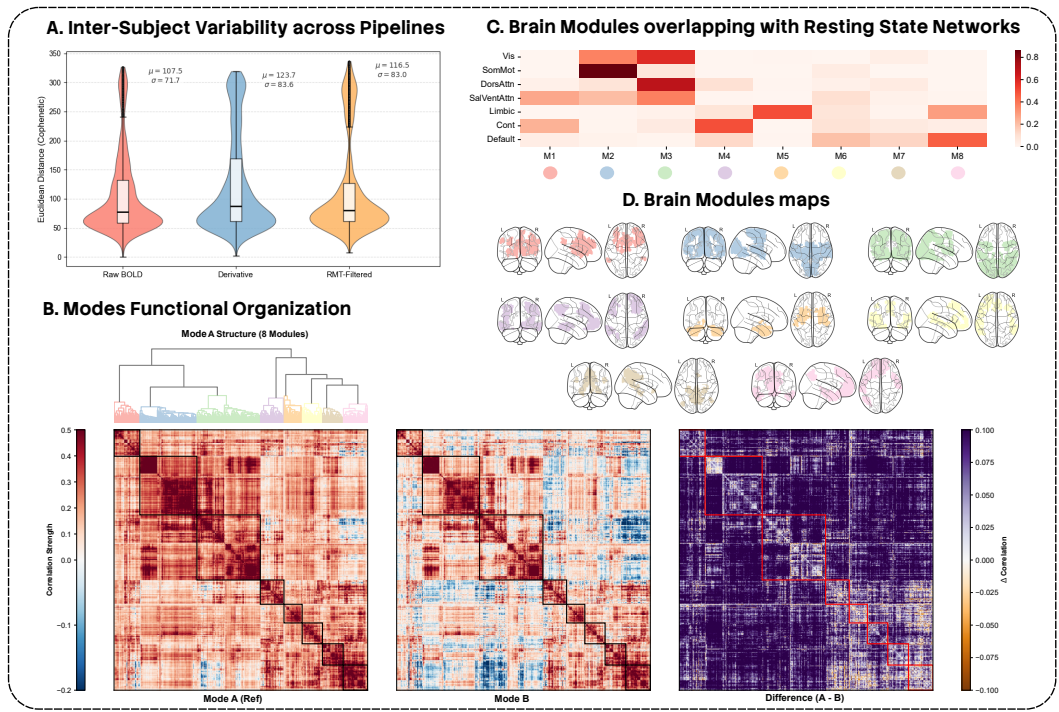

FIG. 4. **LRG–RMT analysis at $N' = 600$.** Population-level functional connectivity structure obtained using the LRG–RMT pipeline at a finer coarse-graining resolution, illustrating reduced but still detectable structured variability. Mode A: $n = 124$; Mode B: $n = 12$.

*Comparison of Mode B functional connectivity pattern at the $N' = 600$ scale.*

Figure 5 compares the Mode B functional connectivity pattern obtained using raw BOLD correlations (Pipeline 1) and the RMT-filtered pipeline (Pipeline 3) at the scale $N' = 600$. Although structured functional patterns remain detectable after RMT filtering, they appear weaker and more fragmented than at intermediate resolutions, consistent with increased noise contamination at higher dimensionality.

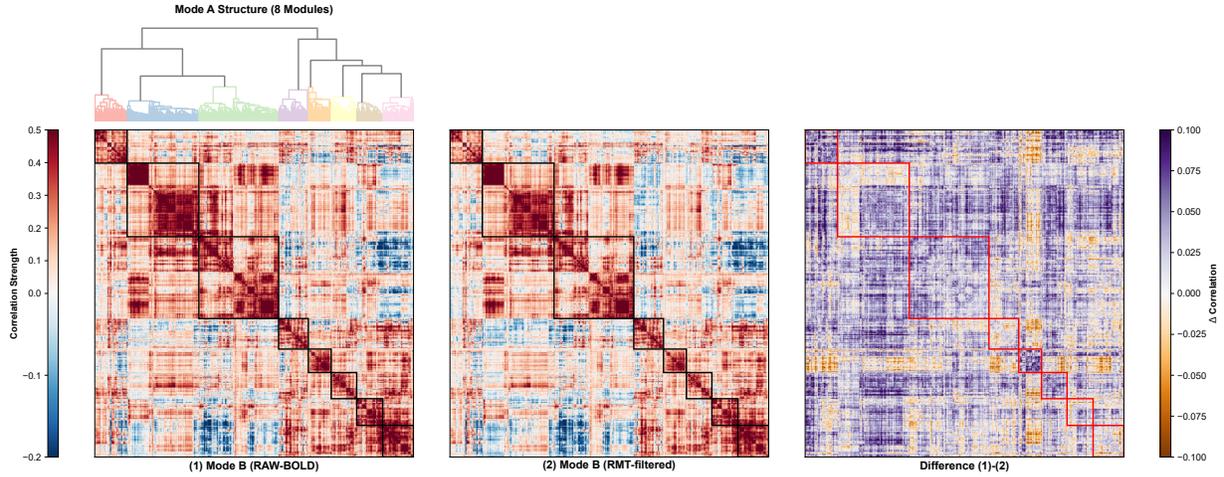

FIG. 5. **Mode B functional organization at $N' = 600$.** Comparison of functional connectivity matrices obtained using raw BOLD correlations (Pipeline 1) and the LRG–RMT pipeline (Pipeline 3), together with their difference.

*Brain modules description at the $N' = 600$ scale.*

Next, we examined the anatomical composition of the LRG-derived modules at the $N' = 600$ resolution. Figure 6 shows that, despite the finer parcellation, modules continue to map onto coherent sets of anatomical regions, indicating that LRG preserves anatomically meaningful organization across scales.

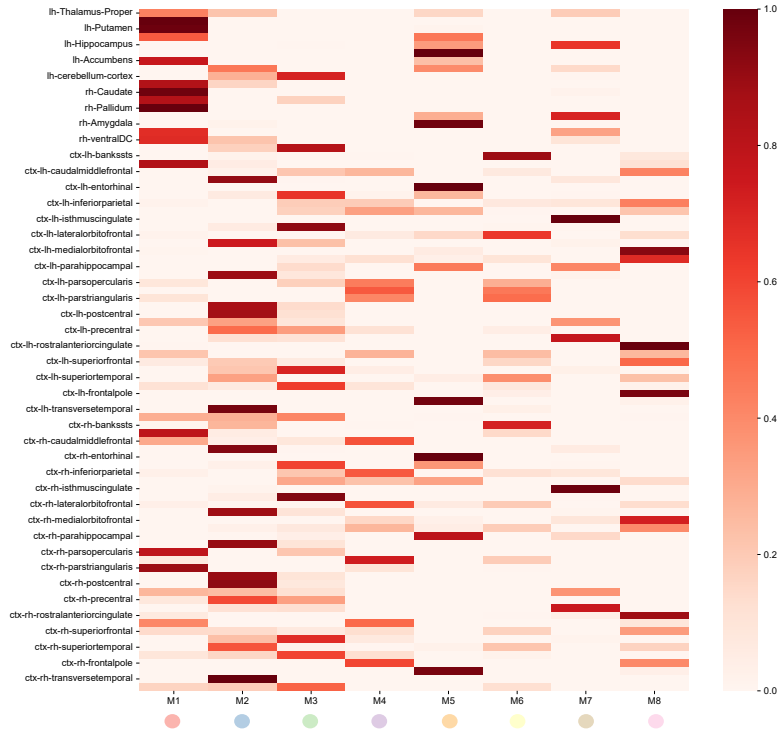

FIG. 6. **Anatomical characterization of LRG-derived modules at $N' = 600$.** Heatmap showing the overlap between data-driven structural modules (columns) and anatomical regions defined by the Desikan–Killiany atlas (rows).

## C. LRG–RMT analysis at the $N' = 60$ scale

Finally, we repeated the LRG–RMT analysis at a coarser resolution ($N' = 60$), where the network is strongly aggregated and individual regions are merged into large-scale units. As shown in Fig. 7, the population-level structure becomes dominated by a single broad mode, indicating a loss of functional differentiation across subjects. At this scale, inter-subject variability is strongly attenuated, and the remaining structure primarily reflects a shared functional backbone rather than distinct organizational regimes.

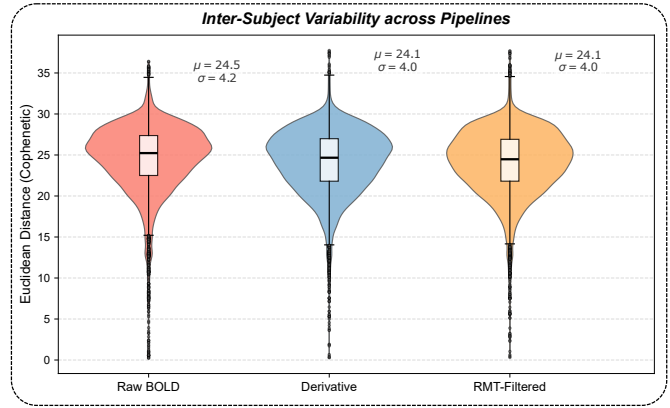

FIG. 7. **LRG–RMT analysis at $N' = 60$.** Population-level functional connectivity structure obtained using the LRG–RMT pipeline at a finer coarse-graining resolution.

Together, these observations reinforce that functional differentiation is maximized at scales where statistical reliability and structural specificity are jointly balanced.


\end{document}